\documentclass[journal=cmatex,manuscript=article]{achemso}

\usepackage{amsmath}
\usepackage{graphicx}

\title{Electrochemical Growth of Full Volume Meissner Effect Superconducting Ba$_{1-x}$K$_x$BiO$_3$}

\author{Oleksandr Foyevtsov}
\affiliation{Department of Physics \& Astronomy, Stewart Blusson Quantum Matter Institute, University of British Columbia, Vancouver, British Columbia, Canada V6T 1Z1}

\author{Kateryna Foyevtsova}
\affiliation{Department of Physics \& Astronomy, Stewart Blusson Quantum Matter Institute, University of British Columbia, Vancouver, British Columbia, Canada V6T 1Z1}

\author{Mohamed Oudah}
\affiliation{Department of Physics \& Astronomy, Stewart Blusson Quantum Matter Institute, University of British Columbia, Vancouver, British Columbia, Canada V6T 1Z1}

\author{Dan Bizzotto}
\affiliation{Department of Chemistry, University of British Columbia, Vancouver, British Columbia, Canada V6T 1Z1}
\email{bizzotto@chem.ubc.ca}

\author{George Sawatzky}
\affiliation{Department of Physics \& Astronomy, Stewart Blusson Quantum Matter Institute, University of British Columbia, Vancouver, British Columbia, Canada V6T 1Z1}


\begin{document}

\begin{abstract}
In this work we show the results of electrochemical synthesis of Ba$_{1-x}$K$_x$BiO$_3$ crystals using three different experimental setup configurations. The setups differ between each other by varying level of control over physical variables and progressing from a two-electrode to a three-electrode configuration using highly oriented platinum counter electrode. By systematic analysis of the temperature dependence of the magnetic susceptibility at the superconducting transition of the collected crystals we observe that an order of magnitude sharper superconducting transition is achieved for the most comprehensive three-electrode setup. In addition, the level of chemical substitution can be controlled by the value of the overpotential versus reference electrode. We demonstrate that with this technique a full volume Meissner effect can be achieved with a sharp transition temperature for the superconducting crystal.
\end{abstract}

\section{Introduction}

Shortly after the discovery of high-temperature superconductivity in cuprates it was also observed in a perovskite insulator BaBiO$_3$ upon chemical substitution \cite{ref1,ref2,ref3}. Such chemical substitution is realized by partial substitution of Ba with K cations in Ba$_{1-x}$K$_x$BiO$_3$ (BKBO) structure. Upon increasing K concentration $x$, BKBO undergoes an insulator to metal transition which coincides with the structural monoclinic to tetragonal transition \cite{ref4}. Further increase of K concentration leads to superconductivity with the highest $T_c$ of about 32 K. Unlike the cuprates the BKBO crystal structure is three-dimensional and its phase diagram has no sign of magnetic ordering. Extensive experimental research shows a number of unconventional features of superconductivity in BKBO, for example, absence of discontinuities in the magnetic susceptibility and specific heat \cite{ref5} at $T_c$ but a sharp transition in electronic transport, complex electron-phonon coupling \cite{ref6}. BKBO also exhibits an unusual electronic structure together with strong electron--lattice coupling and unconventional critical-field behavior. The temperature dependence of the critical field also deviates significantly from the universal behaviour for weak electron-phonon coupling superconductors \cite{ref8}. Due to a number of such unconventional properties the microscopic origin of superconductivity in BKBO is still highly debatable.

The conductivity in metallic BKBO is isotropic, however the temperature dependence varies strongly in the published reports. Varying from metallic to pseudogap like and changing slope behaviour with temperature \cite{ref9,ref10,ref11}. It is not clear if this varying behaviour is due to crystal quality or inhomogeneities. Some reports suggest that the variation in electronic transport along with spectroscopic data can be the result of coexistence of metallic and insulating phases \cite{ref12,ref13,ref14} and/or inhomogeneities of potassium distribution \cite{ref15}. This suggestion is also consistent with the reported small Meissner phase volume \cite{ref11} in the superconducting state, which shows a large scattering among the published reports but values of less than 50\% are very common. Poly and single crystalline BKBO samples can be synthesized with a variety of techniques. While historically it was first synthesized using electrochemical growth, high pressure and high temperature growth \cite{ref16,ref17}, thermal evaporation \cite{ref18}, direct precipitation \cite{ref19,ref20}, floating zone \cite{ref11} as well as post-growth annealing \cite{ref2,ref21} to compensate for oxygen vacancies were explored in order to improve crystal quality. The system crystallizes easily in these techniques. However, the low structural phase purity, which results in a very broad transition width in the magnetic susceptibility, is a downside of the techniques mentioned above. To date the most successful approaches were electrochemical growth \cite{ref22} and post-growth annealing in oxygen flow \cite{ref15,ref21} for which up to an order of magnitude sharper transition widths were reported. The post-growth annealing in oxygen is an effective approach that can largely fill oxygen vacancies. However, the potassium homogeneity is set during the crystal growth. If such inhomogeneity is present, then a part of the crystal volume will not be superconducting and remain as undoped BaBiO$_3$ phase or insufficiently doped BKBO phase \cite{ref15}. Thus, it is highly desirable to achieve targeted stoichiometry and uniform homogeneity during the crystal synthesis.

In this work we demonstrate that high quality BKBO single crystals with sharp superconducting transition and full volume Meissner phase can be systematically grown electrochemically via anodic electrocrystallization \cite{ref23,ref24}.

\[
2\mathrm{BiO}_{3}^{3-} + \mathrm{K}^{+} + \mathrm{Ba}^{2+}
\rightleftharpoons 2(\mathrm{BaK})\mathrm{BiO}_{3} + 3e^{-}
\]

We show in detail how improvement of the electrochemical setup which includes adding an appropriate reference electrode, a symmetric counter electrode, and monitoring the crystal growth in-situ using electrochemical impedance spectroscopy (EIS) influences superconducting properties of the grown BKBO crystals. We believe that revisiting a number of experiments using these electrochemically grown high quality BKBO single crystals may help to address some of the existing controversies in this system.

\begin{figure}[htbp]
  \centering
  \includegraphics[width=0.58\linewidth]{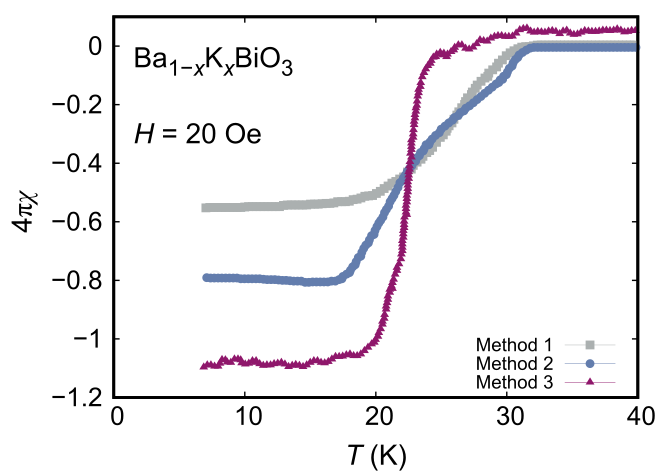}
  \caption{Comparison of magnetic susceptibility for the three successive growth configurations (Methods 1--3) described in the text ($H=20$ Oe).}
  \label{fig:summary}
\end{figure}

Figure~\ref{fig:summary} summarizes the evolution of the magnetic susceptibility across the superconducting transition for crystals grown using the three successive growth configurations (Methods 1--3) investigated in this work. With progressive improvements in thermal and electrochemical control, the transition becomes markedly sharper and the superconducting volume fraction increases, reaching the full-volume Meissner response for the final three-electrode configuration.

\section{Experimental}

Three different setups for electrochemical crystal synthesis, referred to below as Methods 1--3, were used in this work, which are discussed in the next section in detail. For all setups we used 250 ml Teflon crucibles that were loaded in a cylindrical tightly sealed stainless steel chamber as is schematically shown in Fig.~\ref{fig:method1}(a). For all growth experiments as the starting materials we used a mixture of 6 g of Ba(OH$_2$)$\cdot$8$\cdot$H$_2$O (99.9\%), 15.75 g of Bi$_2$O$_3$ (99.9\%), and 210 g of KOH (98\%). The chosen ratio of materials results in chemical substitution corresponding to the highest superconducting transition temperature using setups in Fig.~\ref{fig:method1}(a) and Fig.~\ref{fig:method2}(a). Prior to every growth, the mixed materials were heated to 225$^\circ$ C and homogenized for 7 hours in the sealed growth chamber under flow of humid nitrogen of 10 ccm regulated by a precise two-membrane regulator. A Teflon magnetic stirrer (50 rpm) was used to facilitate homogenization. Nitrogen (5N purity) was bubbled through a water flask that was kept at 55$^\circ$ C first and then driven through the growth chamber. Nominal solution temperatures between 210$^\circ$ C and 270$^\circ$ C were tested, while 225$^\circ$ C resulted in the best crystal quality, so this temperature was used for all experiments presented in this paper. Depending on the setup the crystal growth duration varied between 2 and 10 days.

For the constant current two-electrode setup, 0.2--10 mA was used. High purity platinum wires were used for electrodes, see also discussion for details. For the last, most comprehensive setup shown in Fig.~\ref{fig:method3}(a), a Metrohm Autolab PGSTAT204 potentiostat configured in a three electrode arrangement was used for growth control. The crystal batch used for the magnetic-susceptibility result in Fig.~\ref{fig:method3} was grown at a nominal 50 mV versus a BKBO reference electrode with $x\approx0.4$. The detailed EIS analysis in Fig.~\ref{fig:eisparams} was obtained from a separate growth at nominal 35 mV using the same three-electrode cell geometry and thermal conditions. In this setup the counter electrode was made of 19$\times$3 cm$^2$ oriented (100) polycrystalline platinum foil in a shape of a cylinder positioned around both reference and working electrodes. The working electrode was a 0.5 mm diameter 5N Pt wire flame-polished to form an approximately 1 mm recrystallized, faceted bead at its end. In all experiments the reference and working electrodes were immersed 15mm into the solution at the midpoint of the solution depth.

The temperature dependence of the superconducting transition was measured using a Quantum Design MPMS.

\section{Results and Discussion}

\begin{figure}[htbp]
  \centering
  \includegraphics[width=0.75\linewidth]{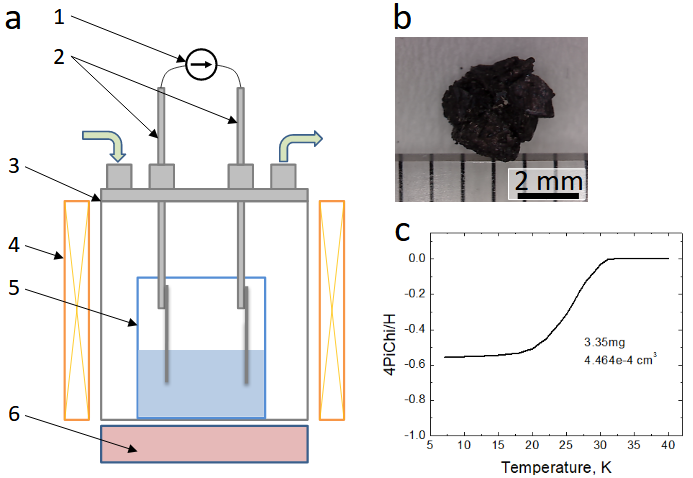}
  \caption{Method 1 -- Simple electrochemical growth setup (a). Here, 1 -- constant current source, 2 -- movable electrodes, 3 -- sealed growth chamber, 4 thermal insulation, 5 -- 320 ml Teflon crucible with growth solution shown in blue, 6 -- spinner and a hot plate. BKBO crystals (b) grown with the setup to the left, and (c) magnetic susceptibility of the crystal above.}
  \label{fig:method1}
\end{figure}

The BKBO system is known to crystallize rather easily using electrochemical growth even with simple control over the growth conditions. However, the poor crystal quality obtained under these conditions requires a closer look at how the growth conditions influence the crystal quality. In our work we start with Method 1, the basic setup shown in Fig.~\ref{fig:method1}(a).

\subsection{Method 1: Constant current growth using a two-electrode configuration}

Here, we use a two-electrode configuration in an air tight steel chamber with 1 cm thick periphery wool insulation and open loop bottom heating control. Both electrodes were untreated 0.5mm Pt wires at about 15mm depth into the solution. Constant electrical current of 2.5 mA was used for growth control. Large crystals with mostly irregular shape, Fig.~\ref{fig:method1}(b), are consistently grown on a positively biased electrode in about two days. The magnetic susceptibility of the collected crystal across the superconducting transition is shown in Fig.~\ref{fig:method1}(c). Here, the onset of superconductivity is observed at around 32 K for the given starting ratio of materials and did not depend on the value of growth electrical current. The transition is smooth and very broad with Meissner phase at the lowest measured temperature reaching only 55 volume percent. The broad character of the transition may be indicative of the presence of multiple structural domains with varying chemical substitution levels and different $T_c$. The monotonic character of this broad transition may suggest the presence of statistically large number of domains. This is also consistent with the small Meissner phase volume such that only about half of the sample volume becomes superconducting. It has to be noted that growth current variation between 500 uA and 5 mA did not significantly modify superconducting transition width or its monotonicity, but larger currents results yielded larger growth rates.

\subsection{Method 2: Constant current growth using an improved two-electrode configuration}

\begin{figure}[htbp]
  \centering
  \includegraphics[width=0.82\linewidth]{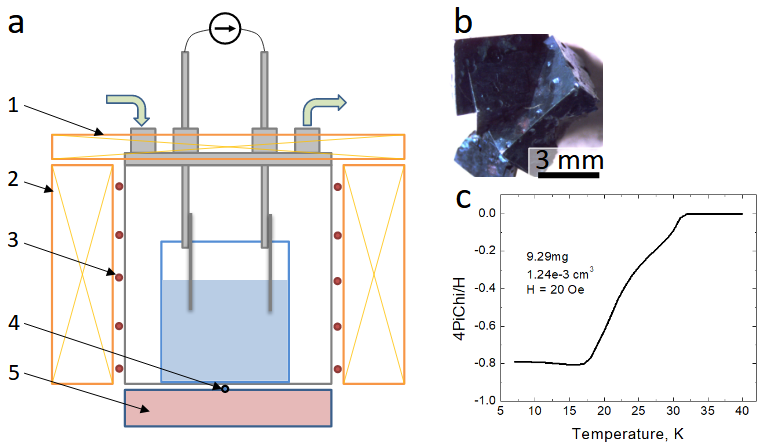}
  \caption{Method 2 -- An improved growth setup (a), here, in addition to the setup in Fig.~\ref{fig:method1}(a) added top thermal insulation (1), increased thickness of the side walls insulation (2), added evenly winded tape heating element (3), feedback thermocouple (4), and only spinner used in (5). BKBO crystals (b) grown with the setup to the left, and (c) magnetic susceptibility of the crystal above.}
  \label{fig:method2}
\end{figure}

In order to improve the crystal quality, we made changes to the growth setup by improving the thermal insulation and switching from the bottom heating source to a circular peripheral heating element embedded into a closed loop regulation circuit with a thermocouple located at the bottom of the steel chamber which is shown schematically in Fig.~\ref{fig:method2}(a). After thorough testing, such modifications resulted in smaller temperature variation across the growth solution at 225$^\circ$ C, decreasing from 20$^\circ$ C for the setup in Fig.~\ref{fig:method1}(a) to under 10$^\circ$ C for the setup in Fig.~\ref{fig:method2}(a). The crystals grown using this improved setup with the same ratio of starting materials as in the previous setup resulted in crystals shown in Fig.~\ref{fig:method2}(b) when using a 2.5 mA constant growth current. The collected crystals are characterized by large cubic-shaped inter-grown domains with distinct circular defects on their surfaces, possibly a result of screw dislocations formed during the growth. Superconducting transition probed by magnetic susceptibility for these crystals is shown in Fig.~\ref{fig:method2}(c). Similarly to the Fig.~\ref{fig:method1}(c), it shows an onset of superconductivity at 32 K and a similar width with a slightly larger Meissner phase volume. However, unlike in the previous experiment the transition is not smooth but features a distinct step-like character. Such character suggests presence of a fewer number of relatively homogeneously doped domains each with a different chemical substitution level, $x$, as compared to the previous experiment.

\subsection{Method 3: Constant potential growth using a three-electrode configuration with in-situ EIS}

In the final and most comprehensive setup in this work the top insulation is further increased and we also added the bottom thermal insulation as shown schematically in Fig.~\ref{fig:method3}(a). The most significant modification was the from a constant current two-electrode setup to a potentiostatic three-electrode setup. The final setup used for growth is shown in Fig.~\ref{fig:method3}(b) and (c) without and with top insulation respectively. Following the pioneering work in \cite{ref22}, a BKBO crystal with $x\approx0.4$, prepared using a constant-current two-electrode configuration, was used as the reference electrode. Also, the counter electrode was changed to be a large oriented (100) polycrystalline platinum foil bent into a cylinder encircling the reference and working electrodes. The 0.5 mm diameter 5N Pt working electrode was flame-polished to form an approximately 1 mm recrystallized, faceted bead at the end in order to provide a reproducible initial surface for growth. Two representative potentiostatic growths are distinguished here: the 50 mV batch, which produced the crystal characterized in Fig.~\ref{fig:method3}(e), and the 35 mV batch, which is used for the detailed EIS analysis below. For the 35 mV dataset analyzed below, in addition to monitoring the dc current, EIS spectra were measured at the deposition potential every 10 min using a 5 mV RMS sine wave potential perturbation over a frequency range of 0.1Hz to 100 kHz. The data was fit to equivalent circuits using dearEIS \cite{ref25}.

\subsection{Crystal Characterization}

\begin{figure}[htbp]
  \centering
  \includegraphics[width=0.96\linewidth]{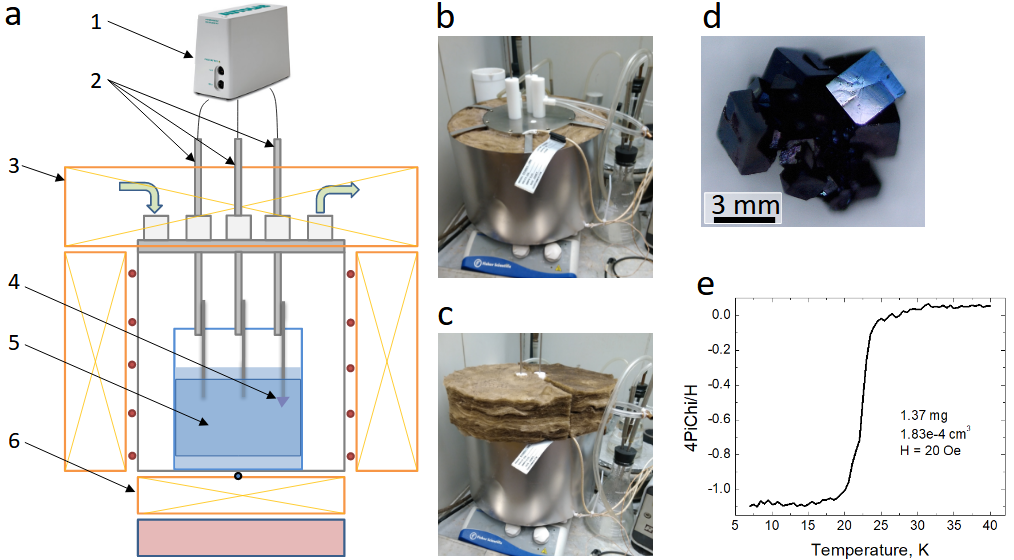}
  \caption{Method 3 -- The final setup used in this work (a), here the additional improvements include a potentiostat (1), three-electrode configuration (2), top thermal insulation with increased thickness (3), reference BKBO electrode (4), large area Pt foil in a shape of a cylinder used as a counter electrode (5), and bottom thermal insulation (6). The actual view of the growth chamber with insulation and electrodes are shown in (b) and (c). BKBO crystals (d) grown with the setup to the left, and (e) magnetic susceptibility of the crystal above.}
  \label{fig:method3}
\end{figure}

The morphology of the grown crystals using this setup, like in the previous setup, also reveals the cubic character of the crystal structure with large inter-grown domains that are shown in Fig.~\ref{fig:method3}(d). However, these crystals show no indication of circular defects and the surfaces are flat and shiny. After separation of domains, we measured the temperature dependence of magnetic susceptibility on these single crystals that is shown in Fig.~\ref{fig:method3}(e). The crystal shown in Fig.~\ref{fig:method3}(e) was separated from the 50 mV aggregate versus the BKBO reference electrode. Several qualitative differences relative to the previous experiments are observed here. First, the transition width is much sharper for these crystals indicating high homogeneity of chemical substitution. Second, to within the experimental error, the sample demonstrates full Meissner phase. It has to be noted that despite the identical ratio of starting materials and growth temperature, the crystal has a lower $T_c$ than those obtained under the previous two-electrode constant-current conditions. This indicates a different superconducting composition or state, although $T_c$ alone does not uniquely determine the potassium content across the BKBO superconducting dome. Earlier electrochemical work established that the applied potential can influence potassium incorporation \cite{ref22}.

\subsection{EIS characterization}

\begin{figure}[htbp]
  \centering
  \includegraphics[width=0.62\linewidth]{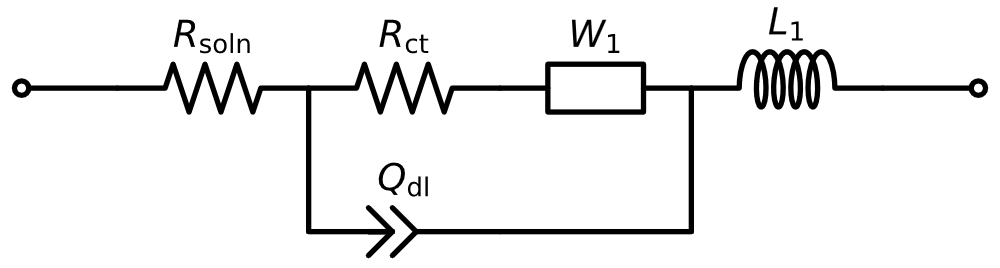}
  \caption{Equivalent circuit model used to fit the EIS data: $R_{\mathrm{soln}}$ -- solution resistance; $Q_{\mathrm{dl}}$ -- constant phase element modeling the double layer capacitance; $R_{\mathrm{ct}}$ -- charge transfer resistance modelling the electrochemical crystal growth kinetics; $W_1$ -- Warburg impedance modelling the mass transport of the reactants; $L_1$ -- inductance modelling the non-ideal connectors and influence of the environment on the EIS measurement.}
  \label{fig:ecm}
\end{figure}

The detailed EIS analysis below was performed on the separate 35 mV growth versus the BKBO reference electrode. EIS was performed during the crystal growth to characterize the changes in the interface during the nucleation and growth of the crystal on a flame annealed Pt bead electrode. The EIS were performed at the growth potential. The current during crystal growth was small (uA) which suggests that the interface was slowly changing which was assumed to be constant over the course of the EIS measurement (a few minutes). Only the data from 5 Hz to 10 kHz was used for fitting to equivalent circuit models (ECM) as it was the most reliable range with stray capacitance affecting the high frequency data and convection/stirring affecting the data below 5Hz. The ECMs used are shown in Fig.~\ref{fig:ecm}.

\begin{figure}[htbp]
  \centering
  \includegraphics[width=0.43\linewidth]{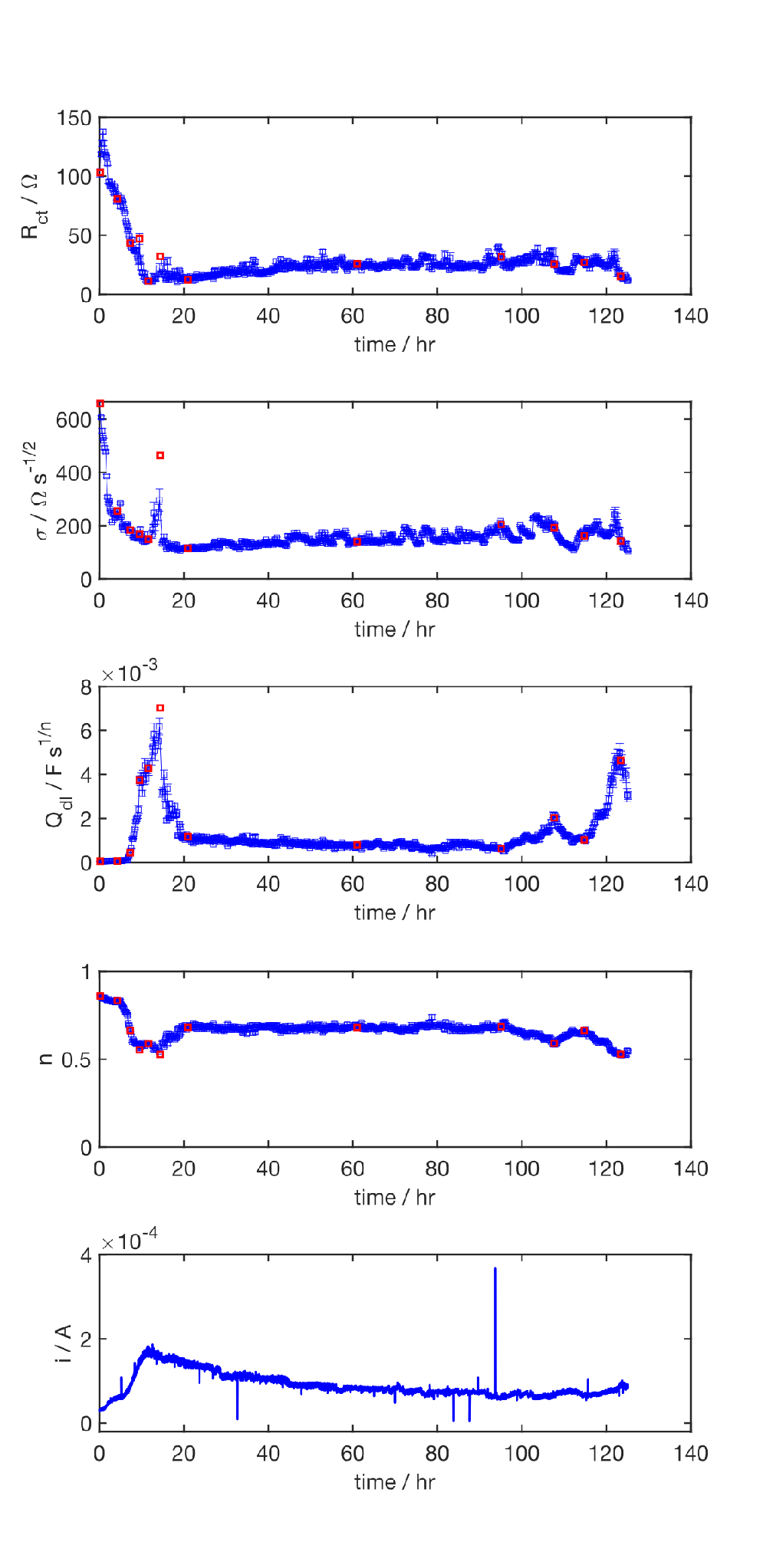}
  \caption{Some of the ECM fitted parameters for the 35 mV batch (a) $R_{\mathrm{ct}}$, b) Warburg impedance ($\sigma$), c) $Q_{\mathrm{dl}}$ and d) $n$ the CPE exponent) determined using the circuit shown in Fig.~\ref{fig:ecm} (the complete fitting results are given in SI) e) is the dc current measured during deposition. The Nyquist and Bode fits are provided in SI for EIS data measured at the red time points.}
  \label{fig:eisparams}
\end{figure}

The ECM fitting results are shown in Fig.~\ref{fig:eisparams} along with the dc current over the 120 h crystal growth process. The $R_{\mathrm{soln}}$ and $L$ (shown in SI) do not change during the crystal growth (1 Ohm and 0.5 uH respectively) confirming the stability of the electrochemical setup.

The parameters from the equivalent circuit fitting are shown in Fig.~\ref{fig:eisparams} with the uncertainty shown as error bars. A selection of the 500 fits are given in SI showing the quality of the fitting. The $\chi^2$ were consistently smaller than $10^{-5}$. The small errors in the parameter estimates shows that the equivalent circuit used did not over parameterize the analysis which suggests that the parameters are all meaningful and can be assigned to physical characteristics of the electrochemical interface.

A constant phase element
\[
Z_{\mathrm{CPE}}=\frac{1}{Q_{\mathrm{dl}}(j2\pi f)^n}
\]
and was used to model the working electrode interface. A CPE can capture the non-ideal capacitive nature of a changing interface where the surface area and dielectric constant evolves non-uniformly with time. In this case, a growing crystal was not observed to be uniformly occurring over the whole Pt working electrode surface necessitating the use of a CPE. This works well for these crystal growth cases where the interface will contain both Pt and Pt coated with growing crystals. The parameter $n$ reflects this non-ideality: $n=1$ represents a perfect capacitor, $n=0$ a perfect resistor, with values in-between characterizing a leaky capacitor or resistor. In this case, $n$ starts at 0.85 and slowly decreases to 0.6, representing an interface which has significant heterogeneity, reflecting the presence of growing crystals. The $Q_{\mathrm{dl}}$ parameter significantly changes during the initial stages of crystal growth at short times ($<20$hrs). This is in accordance with the increase in the dc current and a decrease in the $R_{\mathrm{ct}}$. These dramatic changes can be associated with an increase in the area of the interface (since the crystals are conductive given that they support electrochemical growth) and a change in the interfacial composition. It appears that the initial 8 hours is constant with a dramatic increase in $Q_{\mathrm{dl}}$ observed at 10 hours. This corresponds to an increase in the crystal growth current. These dramatic changes occurring at 8 h can be explained by the growth of the crystal on the Pt surface after a long period of nucleation sites being created until some are stable enough to grow. The rapid growth at 10 hours increases the magnitude of $Q_{\mathrm{dl}}$ by 25X which is most reasonably interpreted as a fast growth from a few nuceli. This is also observed as a decrease in the CPE exponent ($n$) which can be interpreted as a rougher surface resulting in a larger surface heterogeneity. The dc current also reflects increased rate of crystal growth (see electrochemical reaction). The resistance to charge transfer ($R_{\mathrm{ct}}$), or rate of electrochemical crystal growth decreases significantly over the first 10 h. This can also be explained as initial creation of nuceli on a Pt surface which can be a kinetically slow process. But after the formation of nuclei, growth on existing BKBO crystals is more facile which is reflected in a decrease in the $R_{\mathrm{ct}}$ or increase in the rate (coincident with the increase in dc current and surface area (via $Q_{\mathrm{dl}}$)). The Warburg impedance which characterizes the mass transport to the surface and is sensitive to the surface area also shows a lower impedance when the surface area increases due to facile mass transport to a surface with increased crystal growth and larger area. Overall, the ECM parameters are consistently describing the nucleation and growth of crystals on some parts of the Pt bead electrode.

After 20 h, the growth occurs without significant changes in the ECM parameters and a slow decrease in the dc current. This can be explained as a consistent growth of the BKBO crystal or crystals on the Pt electrode. This occurs up to 100 h where significant changes are observed in the $Q_{\mathrm{dl}}$. No significant change was observed in the dc current nor other ECM parameters. $Q_{\mathrm{dl}}$ is sensitive to the heterogeneity of the interface suggesting that this increase maybe a signature of the start of a new crystal growth process occurring on another part of the Pt surface, or an extra phase within the growing crystal. The changes are similar to the initial crystal growth parameters which suggests the presence of another growing crystal on a different part of the Pt electrode. This observation can be useful to ensure that the crystal grown is of high quality, as it is important to know when to stop the crystal growth to ensure high quality samples without significant overgrowth. Monitoring the in-situ EIS can be a strategy for deciding when the growth should be stopped.

The corresponding 50 mV growth was analyzed in the same manner and the results are summarized in the Supplementary Information. Compared with the 35 mV growth, the deposition current is more than an order of magnitude larger, while $R_{\mathrm{ct}}$ and the Warburg parameter ($W$) are substantially smaller and $Q_{\mathrm{dl}}$ is approximately an order of magnitude larger. The 50 mV data retain the same broad evolution from an induction stage into increasing crystal growth, but the much larger deposition rate maintains a large electrochemically active interface as growth proceeds. The mass-transport contribution does not indicate a simple transition to strongly diffusion-limited growth. Thus, a change of only 15 mV produces a pronounced change in the electrocrystallization rate, illustrating the importance of potentiostatic control and the additional information available from in-situ EIS informing on the crystal growth process in-situ.

\section{Conclusions}

By systematic study of the importance of the control of a variety of growth conditions in the electrochemical synthesis of BKBO crystals, we find that the highest crystal quality can be achieved with potentiostatic growth using a three-electrode setup and advanced temperature control. Comparison of the 35 and 50 mV potentiostatic growths shows that a 15 mV change in controlled working-electrode potential produces a large change in deposition rate and in the fitted EIS parameters. Due to the direct control over the potential difference in three-electrode electrochemcial synthesis, this approach may also be relevant to other transition metal oxide families where indirect physical quantities are tuned during synthesis, such as pressure, or initial ratio of components. Moreover, the ability to monitor the growth using in-situ EIS measurements provides an opportunity to carefully choose the conditions where growth can be terminated in order to ensure high quality crystals are recovered

\section{Acknowledgement}

We would like to thank David Brown for the kind help in providing us facilities for this research in his lab. This work was supported by NSERC, CIFAR, and the Max Planck -- UBC Stewart Blusson Quantum Matter Institute.


\begin{thebibliography}{25}

\bibitem{ref1} Cava RJ, Batlogg B, Krajewski JJ, Farrow R, Rupp LJ, White AE, Short K, Peck WF, Kometani T. Superconductivity near 30 K without copper: the Ba$_{0.6}$K$_{0.4}$BiO$_3$ perovskite. Nature. 1988 Apr;332(6167):814--6.

\bibitem{ref2} Hinks DG, Dabrowski B, Jorgensen JD, Mitchell AW, Richards DR, Pei S, Shi D. Synthesis, structure and superconductivity in the Ba$_{1-x}$K$_x$BiO$_{3-y}$ system. Nature. 1988 Jun;333(6176):836--8.

\bibitem{ref3} Mattheiss LF, Gyorgy EM, Johnson Jr DW. Superconductivity above 20 K in the Ba-K-Bi-O system. Physical Review B. 1988 Mar 1;37(7):3745.

\bibitem{ref4} Pei S, Jorgensen JD, Dabrowski B, Hinks DG, Richards DR, Mitchell AW, Newsam JM, Sinha SK, Vaknin D, Jacobson AJ. Structural phase diagram of the Ba$_{1-x}$K$_x$BiO$_3$ system. Physical Review B. 1990 Mar 1;41(7):4126.

\bibitem{ref5} Kumar P, Hall D, Goodrich RG. Thermodynamics of the superconducting phase transition in Ba$_{0.6}$K$_{0.4}$BiO$_3$. Physical Review Letters. 1999 May 31;82(22):4532.

\bibitem{ref6} Meregalli V, Savrasov SY. Electron-phonon coupling and properties of doped BaBiO$_3$. Physical Review B. 1998 Jun 1;57(22):14453.

\bibitem{ref7} Batlogg B, Cava RJ, Rupp Jr LW, Mujsce AM, Krajewski JJ, Remeika JP, Peck Jr WF, Cooper AS, Espinosa GP. Density of states and isotope effect in BiO superconductors: Evidence for nonphonon mechanism. Physical Review Letters. 1988 Oct 3;61(14):1670.

\bibitem{ref8} Hall D, Goodrich RG, Grenier CG, Kumar P, Chaparala M, Norton ML. Magnetization measurements on single crystals of superconducting Ba$_{0.6}$K$_{0.4}$BiO$_3$. Philosophical Magazine B. 2000 Jan 1;80(1):61--79.

\bibitem{ref9} Naamneh M, Yao M, Jandke J, Ma J, Risti\'c Z, Teyssier J, Stucky A, van der Marel D, Gawryluk DJ, Shang T, Medarde M. Cooling a polaronic liquid: Phase mixture and pseudogap-like spectra in superconducting Ba$_{1-x}$K$_x$BiO$_3$. arXiv preprint arXiv:1808.06135. 2018 Aug 18.

\bibitem{ref10} Sleight AW. Bismuthates: BaBiO$_3$ and related superconducting phases. Physica C: Superconductivity and its Applications. 2015 Jul 15;514:152--65.

\bibitem{ref11} Noji T, Kato T, Imai Y, Koike Y. Single-Crystal Growth of the Superconducting Ba$_{1-x}$K$_x$BiO$_3$ by the Floating-Zone Method. In AIP Conference Proceedings 2006 Sep 7 (Vol. 850, No. 1, pp. 671--672). American Institute of Physics.

\bibitem{ref12} Nagata Y, Mishiro A, Uchida T, Ohtsuka M, Samata H. Normal-state transport properties of Ba$_{1-x}$K$_x$BiO$_3$ crystals. Journal of Physics and Chemistry of Solids. 1999 Dec 1;60(12):1933--42.

\bibitem{ref13} Tajima S, Uchida S, Masaki A, Takagi H, Kitazawa K, Tanaka S, Katsui A. Optical study of the metal-semiconductor transition in BaPb$_{1-x}$Bi$_x$O$_3$. Physical Review B. 1985 Nov 15;32(10):6302.

\bibitem{ref14} Chainani A, Yokoya T, Kiss T, Shin S, Nishio T, Uwe H. Electron-phonon coupling induced pseudogap and the superconducting transition in Ba$_{0.67}$K$_{0.33}$BiO$_3$. Physical Review B. 2001 Oct 18;64(18):180509.

\bibitem{ref15} Kim BJ, Kim YC, Kim HT, Kang KY, Lee JM. EXAFS observation of two distinct Bi--O distances below $T_c$ for a Ba$_{0.6}$K$_{0.4}$BiO$_3$ single crystal. Physica C: Superconductivity. 2003 Oct 1;392:286--90.

\bibitem{ref16} Kim DC, Baranov AN, Kim JS, Kang HR, Kim BJ, Kim YC, Pshirkov JS, Antipov EV, Park YW. Superconductivity of Ba$_{1-x}$K$_x$BiO$_3$ ($0.35<x<1$) Synthesized by the High Pressure and High Temperature Technique. Journal of Superconductivity. 2002 Oct;15(5):331--4.

\bibitem{ref17} Kim DC, Baranov AN, Kim JS, Kang HR, Kim BJ, Kim YC, Pshirkov JS, Antipov EV, Park YW. High pressure synthesis and superconductivity of Ba$_{1-x}$K$_x$BiO$_3$ ($0.35<x<1$). Physica C: Superconductivity. 2003 Jan 1;383(4):343--53.

\bibitem{ref18} Utz B, Wiest F, Prusseit W, Berberich P, Kinder H. Ba$_{1-x}$K$_x$BiO$_3$ epitaxy on various substrate materials. IEEE Transactions on Applied Superconductivity. 1995 Jun;5(2):1351--4.

\bibitem{ref19} Liu SF, Fu WT. Synthesis of superconducting Ba$_{1-x}$K$_x$BiO$_3$ by a modified molten salt process. Materials Research Bulletin. 2001 May 1;36(7-8):1505--12.

\bibitem{ref20} Zhao LZ, Yin B, Zhang JB, Li JW, Xu CY, Xie M, Liu SH. Synthesis of BaKBiO superconductor at 260$^\circ$ C by direct precipitation from KOH melts. Physica C: Superconductivity. 1997 Aug 1;282:723--4.

\bibitem{ref21} Mosley WD, Liu JZ, Matsushita A, Lee YP, Klavins P, Shelton RN. Preparation and superconductivity of Ba$_{1-x}$K$_x$BiO$_3$ single crystals. Journal of Crystal Growth. 1993 Mar;128(1-4):804--7.

\bibitem{ref22} Han PD, Chang L, Payne DA. Top-seeded growth of superconducting (Ba$_{1-x}$K$_x$)BiO$_3$ crystals by an electrochemical method. Journal of Crystal Growth. 1993 Mar 1;128(1-4):798--803.

\bibitem{ref23} Norton ML, Tang HY. Superconductivity at 32 K in electrocrystallized barium potassium bismuth oxide. Chemistry of Materials. 1991;3(3):431--434. doi:10.1021/cm00015a015.

\bibitem{ref24} Zheng XG, Taira M, Suzuki M, Xu CN. Growing a periodic microstructure on the superconductor crystal surface by electrocrystallization. Applied Physics Letters. 1998;72(10):1155--1157. doi:10.1063/1.120999.

\bibitem{ref25} Yrj\"an\"a, V. (2022). DearEIS - A GUI program for analyzing impedance spectra. Journal of Open Source Software, 7(80), 4808.

\end{thebibliography}
\end{document}